\documentclass[a4paper, oneside, twocolumn, notitlepage, 10pt]{extarticle_ecoc}
\usepackage{ecoc}

\usepackage{biblatex}
\begin{document}
\selectlanguage{english}    


\title{Squeezed Light Coexistence with  Classical Communication over\\    
10 km Optical Fiber} 


\author{
    Adnan A.E. Hajomer\textsuperscript{*}, Huy Q. Nguyen\textsuperscript{$\dagger$},
    Melis Pahal{\i}, Ulrik L. Andersen, Tobias Gehring
}

\maketitle                  


\begin{strip}
 \begin{author_descr}

    Center for Macroscopic Quantum States (bigQ), Department of Physics, Technical University of Denmark, 2800 Kongens Lyngby, Denmark,
   \textsuperscript{*}\textcolor{blue}{\uline{ aaeha@dtu.dk}} 

  \textsuperscript{*,$\dagger$} \textit{These authors are contributed equally as first authors}


 \end{author_descr}
\end{strip}

\setstretch{1.1}
\renewcommand\footnotemark{}
\renewcommand\footnoterule{}
\let\thefootnote\relax\footnotetext{978-1-6654-3868-1/21/\$31.00 \textcopyright 2021 IEEE}


\begin{strip}
  \begin{ecoc_abstract}
    We report the first coexistence experiment of 1550 nm single-mode squeezed states of light
with a 1310 nm classical telecom channel over a 10 km fiber channel while measuring squeezing using
a locally generated local oscillator. This is achieved using real-time optical heterodyne phase locking,
allowing us to measure up to 0.5 dB of squeezing with a phase noise of $2.2~^{\circ}$.
  \end{ecoc_abstract}
\end{strip}


\section{Introduction}

\begin{figure*}[t]
   \centering
    \includegraphics[width= \linewidth, keepaspectratio]{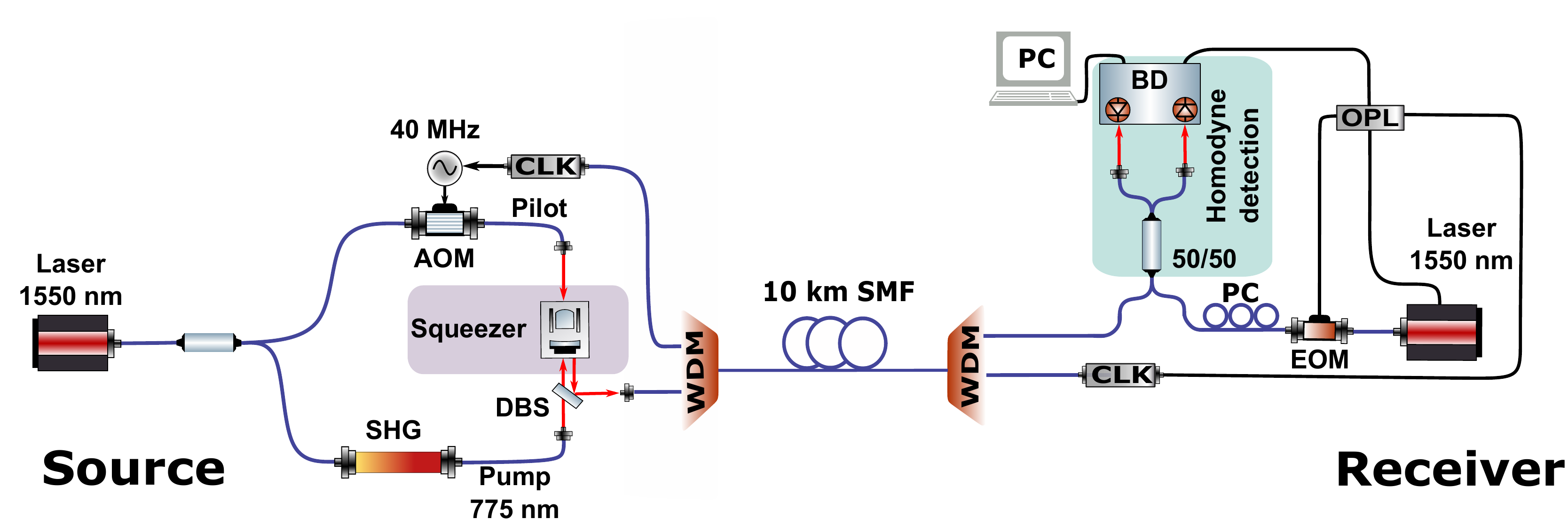}
    \vspace{0.3cm}
    \caption{Experimental setup for coexistence of single-mode squeezing and classical signal. AOM: acousto-optic modulator; SHG: second-harmonic generator module; CLK: 10 MHz reference clock; EOM: electro-optic modulator; DBS: dichroic beam splitter PC: polarization controller; OPL: optical phase locking; BD: balanced detector.}
    \label{fig1}
    \vspace{1cm}
\end{figure*}

Squeezed quantum states of light are non-classical states that exhibit reduced quantum uncertainty in one quadrature, known as squeezing, compared to that of the vacuum state~\supercite{caves1981quantum,yuen1976two}. This unique quantum property has enabled a wide range of applications, including quantum-enhanced measurement of gravitational waves~\supercite{aasi2013enhanced}, measurement-based quantum computing~\supercite{larsen2019deterministic, larsen2021deterministic}, Gaussian boson sampling~\supercite{arrazola2021quantum,zhong2021phase}, and quantum error correction~\cite{menicucci2014fault}. Furthermore, squeezed light also has applications in quantum key distribution (QKD), a well-known application of quantum cryptography, to improve the system's performance. In particular, continuous variable QKD based on squeezed states is highly tolerant to excess noise, channel loss, and imperfect error correction compared to coherent state-based protocols~\supercite{madsen2012continuous, gehring2015implementation, jacobsen2018complete}. Given the practical importance of these applications, the ability to transmit squeezed light over existing fiber networks is crucial for large-scale deployment. 

Several experiments on squeezed light fiber transmission have recently been reported~\supercite{suleiman202240, huo2018deterministic, liu2018long}. However, these experiments typically require dark fibers, which increase costs and limit availability. Therefore, the coexistence of squeezed light and classical telecom channels in the same fiber is desirable for practical applications. In addition, practical aspects such as clock synchronization between  the source and remote receiver and the use of a 
 real local oscillator (LO) scheme for homodyne detection~\supercite{suleiman202240} to measure squeezed light must be considered for real-world applications. 


The coexistence of squeezed light and classical telecom channels presents a significant challenge due to the high optical power of the classical channels and the noise intolerance of squeezed light. To separate the squeezed light from classical channels, the wavelength division multiplexing (WDM) technique can be utilized. However, the challenge of multiplexing squeezed light and classical telecom channels on the same fiber using WDM arises from additional noise generated from bright classical light due to cross-talk and optical nonlinear effects. Therefore, managing fiber non-linearity and cross-talk represents the main challenges for the coexistence of squeezed light with classical channels.

Here, we report the first experiment (to our knowledge) of coexisting single-mode squeezed light and classical light over a 10 km fiber channel, using a local (or real) LO (LLO) for measuring squeezing. This is made possible by maintaining a large wavelength separation between squeezed light  (C-band) and the telecom channel (O-band) so that the excess noise due to fiber non-linearity and cross-talk is negligible. 
Moreover, we have implemented real-time optical heterodyne phase locking~\supercite{suleiman202240} to accurately control the phase of the LLO, achieving a relatively small phase uncertainty of $2.2^{\circ}$. With this setup, we were able to measure up to 0.5 dB of squeezing. This marks a significant achievement in building a quantum network based on continuous variable quantum systems.

 \begin{figure}[t]
   \centering
    \includegraphics[width= \linewidth, keepaspectratio]{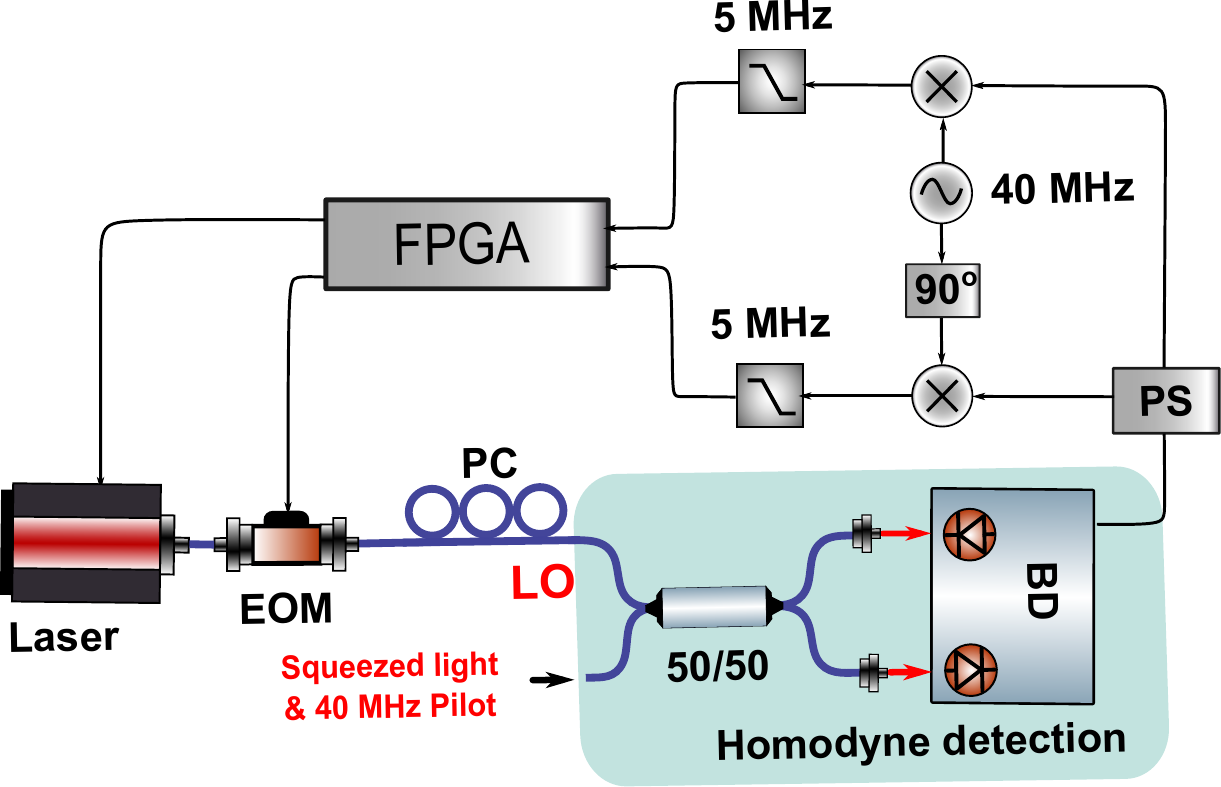}
    \vspace{0.3cm}
    \caption{Phase locking system. PS: power splitter; EOM: electro-optic modulator; PC:
polarization controller; FPGA: field programmable gate array.}
    \label{fig2}
\end{figure} 

 \begin{figure*}[t]
   \centering
    \includegraphics[width= \linewidth, keepaspectratio]{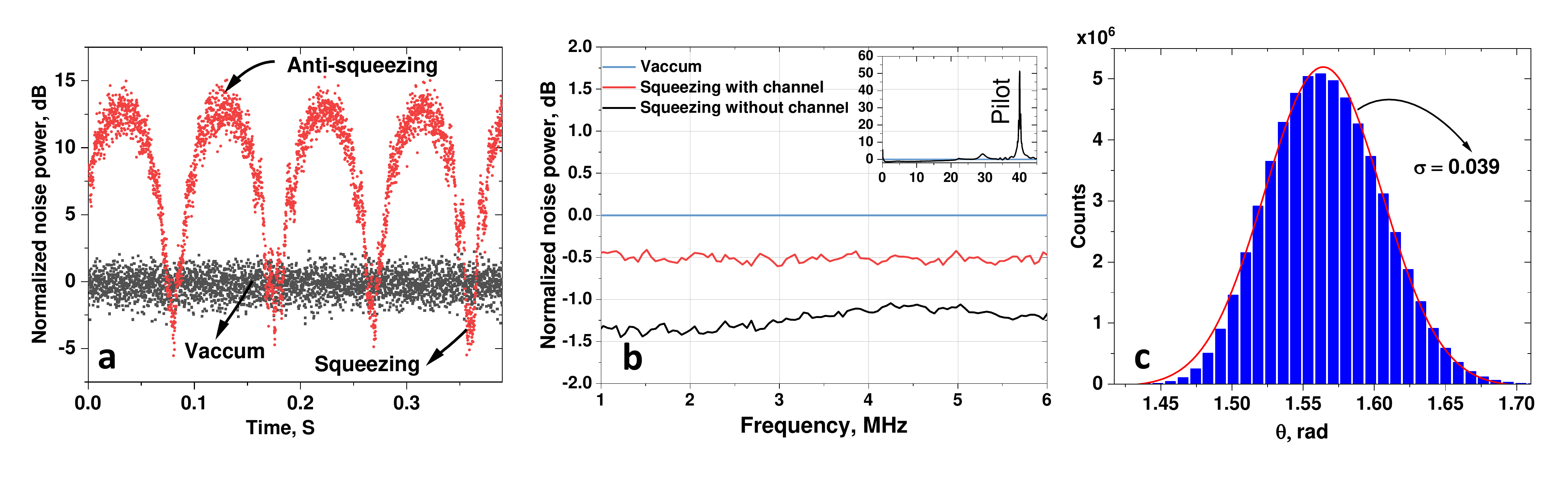}
    \caption{Experimental results. (a) Squeezing measurement using a shared laser. (b) Coexistence with and without a 10 km fiber channel and LLO scheme. (C) The histogram of pilot tone phase profile after phase locking.}
    \vspace{1cm}
    \label{fig3}
\end{figure*}

\section{Experimental setup}

 Figure~\ref{fig1} shows the schematic of the experimental setup used to demonstrate the coexistence of squeezed light and a classical telecom channel. At the source, a 1550 nm continuous wave (CW) laser (NKT Koheras BASIK) with a linewidth of 100 Hz was used as an optical source. Squeezed vacuum  at 1550 nm was generated via parametric down-conversion using an optical parametric oscillator (OPO), pumped with 775 nm light. The OPO is a hemilithic cavity constructed from a planar-convex periodically poled potassium titanyl phosphate (PPKTP) crystal as the nonlinear medium along with a partially reflective mirror. The 775 nm pump field was generated using a commercial second harmonic generator (SHG) waveguide module. The phase locking 
 system of the source was implemented in two stages. First, the squeezed light source’s cavity was stabilized using a Pound-Drever-Hall~\supercite{10.1119/1.1286663} lock in the pump path. Second, a coherent-locking scheme was used to lock the relative phase between the pump field and a pilot tone generated at 40 MHz using a fiber acousto-optic modulator (AOM). The phase-locked pilot tone was then transmitted with squeezed light to the receiver.  More details about the squeezer and source's locking system can be found in~\supercite{arnbak2019compact}.

Synchronizing clocks between the source and the remote receiver is a fundamental requirement for performing a squeezing measurement. This was achieved using a 10 MHz master clock at the source, which was then converted into an optical signal of 1 mW power at 1330 nm using a commercial electrical-to-optical converter from Highland Technology.  The  1330 nm optical clock was wavelength-multiplexed with the 1550 nm single-mode squeezed light using WDM. The multiplexed signals propagated through 10 km standard single-mode fiber (SMF) with a total loss of $\approx 1.8$~dB. At the receiver side, another WDM system was used to demultiplex the squeezed light and the optical clock. The optical clock was then converted to the electrical domain using an optical-to-electrical converter module and sent to the optical phase locking (OPL) system. 

 To measure squeezing, the laser used to generate the LO for balanced homodyne detection must be frequency locked to the source's laser. This is commonly achieved using a transmitted LO (TLO) scheme~\supercite{chapman2023two}, where the same laser is shared between the source and the receiver. However, this arrangement can lead to deteriorating non-linear optical
effects due to the high power of the TLO. Hence, in our system, an independent laser was used to generate the LO at the receiver side and phase-locked to the squeezed light using a real-time optical heterodyne phase-locking system. 

Figure~\ref{fig2} illustrates the schematic of our optical heterodyne phase locking system~\supercite{suleiman202240}. The squeezed vacuum and 40 MHz pilot tone were combined with the LO on a balanced beam splitter, generating a beat signal that indicates the relative frequency and phase difference with respect to the LO, as the 40 MHz pilot tone was phase-locked to the squeezed light at the source. The beat signal was detected using a balanced detector (BD) and demodulated using a 40 MHz function generator that was clock synchronized to the source's function generator using the 1330 nm optical clock.  After the I-Q demodulation process, error signals were generated using a field programmable gate array (FPGA), where a proportional-integral (PI) controller was implemented to drive a piezoelectric wavelength modulator inside the laser for frequency lock, and an electro-optic phase modulator (EOM) for phase locking. After phase locking, the polarization of LO was optimized using a polarization controller (PC). Finally, the output of the BD was digitized using an analog-to-digital card and recorded for later processing.

\section{Results}
We evaluated the performance of our system by measuring squeezing using three different configurations. The first configuration used a TLO that originated from the same laser used for squeezed light generation and shared with the receiver as a LO using another fiber. This setup enabled us to avoid the phase noise of the LLO, making it a reference measurement. Figure~\ref{fig3} (a) shows the normalized noise power of squeezing and anti-squeezing quadratures at 4 MHz, while slowly scanning the phase of the TLO. Notably, we obtained $\approx 3.5$ dB of vacuum noise suppression.

Next, we used the LLO to measure squeezing in case of coexistence with a back-to-back configuration, where the source and the receiver were connected through a short fiber channel. Fig.~\ref{fig3} (b) shows the results of this experiment, with the black line indicating $\approx 1.3$ dB of squeezing over a 5 MHz bandwidth. Compared to the TLO the reduction of squeezing can be attributed to the insertion loss of the WDM and fiber connectors and the residual phase noise from the locking system. 

As phase noise degrades the squeezing level, we measured the phase noise of our locking system. This was done by extracting the phase profile of the 40 MHz pilot tone, shown in the inset of Fig.~\ref{fig3} (b). The resulting histogram of the phase profile is shown in Fig.\ref{fig3} (c), indicating a Gaussian distribution with a relatively small standard deviation of 0.039 rad (corresponding to $2.2^\circ$). This small deviation indicates that the phase uncertainty of our locking system is low, minimizing the degradation of squeezing due to phase noise.

Finally, we evaluated the performance of our system for 10 km fiber transmission.  As shown in Fig.~\ref{fig3} (b), we measured 0.5 dB of squeezing as a result of the loss of the fiber channel and WDM system. This shows the feasibility of coexisting classical channels and squeezed light in the same fiber channel.      

\section{Conclusions}
 We reported the first practical experiment demonstrating the coexistence of single-mode squeezed light and a classical telecom channel using the LLO scheme for squeezing measurement. Moreover, we considered the actual clock synchronization between the source and the remote receiver. Our system relies on real-time optical phase locking and sufficient channel spacing between squeezed light and the classical light. The results lay the foundation for the realization of realistic continuous variable quantum networks where squeezed light needs to propagate to various physically separated network nodes.   

\vspace{0.6cm}
\noindent \textbf{Acknowledgements}

\noindent The authors acknowledge funding from the Carlsberg Foundation (project CF21-0466), Danmarks Frie Forskningsfond (project 0171-00055B), Center for Macroscopic Quantum States (bigQ, DNRF142), and QuantERA II Programme (project CVSTAR) that has received funding from the European Union’s Horizon 2020 research and innovation programme under Grant Agreement No 101017733.


\printbibliography
\end{document}